\documentclass[epj]{svjour}
\usepackage{graphicx,wrapfig,lipsum}
\usepackage{amssymb,amsmath,bbold}
\usepackage{url,hyperref,doi} 
\usepackage{mathtools} 		
\usepackage{float}	 
\usepackage{color}

\makeatletter
\g@addto@macro{\UrlBreaks}{\UrlOrds}
\makeatother

\newcommand{\ie}{{\it i.e.\/}}

\newcommand{\aka}{{\it aka\/}}

\newcommand{\loccit}{{\it loc.cit.\/}}

\newcommand{\nc}{\newcommand} 
\nc{\req}[1]{Eq.\,(\ref{#1})} \nc{\reqp}[1]{Eq.\,(\ref{#1}) on page \pageref{#1}} 
\nc{\rf}[1]{Fig.~\ref{#1}} \nc{\rfp}[1]{Fig.~\ref{#1} on page \pageref{#1}} 
\nc{\rt}[1]{table~\ref{#1}} \nc{\rtp}[1]{table~\ref{#1} on page \pageref{#1}} 
\nc{\pp}{\ensuremath{pp\ }}
\nc{\pA}{\ensuremath{p A\ }}
\nc{\hAA}{\ensuremath{AA\ }}
\nc{\D}{\mathrm{d}}
\nc{\E}{\mathrm{e}}

\begin{document}
\title{Measurement of the Lorentz-Fitz\allowbreak Gerald Body Contraction}
\subtitle{Dedicated to Walter Greiner; October 1935 -- October 2016}
\author{Johann Rafelski\inst{1}\inst{2}}
%
%
\institute{Department of Physics, The University of Arizona 
Tucson, Arizona, 85721, USA}
\date{Submitted: June 25, 2017 / Revised: August 16, 2017 / Print date: \today}
%
\abstract{%
A complete foundational discussion of acceleration in context of Special Relativity (SR) is presented. Acceleration allows the measurement of a Lorentz-Fitz\allowbreak Gerald body contraction created. It is argued that in the back scattering of a probing laser beam from a relativistic flying electron cloud mirror generated by an ultra-intense laser pulse, a first measurement of a Lorentz-Fitz\allowbreak Gerald body contraction is feasible.
\PACS{%
 {03.30.+p}{Special Relativity} 
 } 
} 
\maketitle
%
{\bf Introduction.} ---
Within the relativity framework created by Einstein in 1905~\cite{EinsteinSR}, there is no acceleration. Einstein considered only inertially moving bodies and observers. Imposing Galileo's relativity principle, homogeneity and iso\-tropy of space, and the constancy of the speed of light, Einstein obtained all his results using Lorentz coordinate transformations that follow from these principles. To obtain the relativistic Doppler effect maintaining the relativity principle, Einstein postulated that the light wave phase is a Lorentz invariant. Calling the theory of relativity \lq Special Relativity\rq\  (SR), we consider the consequences of principles here stated along with the effect of acceleration as we introduce, excluding gravity which is referred by the  acronym GR -- we side with authors who take GR to mean \lq Gravity (Relativity)\rq. 
 
Lorentz considers forces and acceleration and relies on acceleration to  transfer a body between different inertial frames of reference. Einstein in 1911 notes that both approaches to relativity are equivalent~\cite{AEEhrenfest}: \lq\lq The claim \ldots of a difference between Lorentz's view and that of mine with regard to physical properties (of body contraction, JR) is not correct. \ldots For a comoving observer it (body contraction, JR) is not present and as such it is not observable; however it is real and in principle observable by physical means by any non-comoving observer.\rq\rq

This situation is described in 1960 by Wolfgang A. Rindler~\cite{Rindler1960}: \lq\lq Relativity offers no detailed explanation in terms of cohesive forces or the like [{\it however, compare~\cite{Rafelski:2017hyt}, Chapter 10 pp. Discussion IV-2)}], yet it predicts the contraction phenomenon as inevitable. This is comparable to some of the predictions based on the energy principle. It must be stressed that the phenomenon is not to be regarded as illusory \ldots it is real in every possible sense.\rq\rq\ The last comments echo the remarks of Einstein of 1911 \loccit. Lajos J\' anossy in his 1971 book calls Lorentz approach \lq physical reality\rq\cite{Janossy71}.

In 1976 John S. Bell, of quantum Bell-inequality fame, presented \lq\lq How to teach special relativity\rq\rq\cite{JSBell}. Bell advances what we call the \lq\lq Lorentz-Bell pedagogy\rq\rq. It relies on acceleration, as small and negligible as need be, but non-zero, allowing a dynamical transfer of a body between different inertial frames of reference, \ie\ a \lq dynamical viewpoint\rq\cite{Nelson15}. Bell in his letter to the author in 1985 worried~\cite{LetterlBook}: \lq\lq Einstein's approach is perfectly sound, and very elegant and powerful (but pedagogically dangerous, in my opinion).\rq\rq\ Bell does not see a difference between his/Lorentz approach and that of Einstein worth mentioning. Furthermore, nothing in his two papers on SR~\cite{JSBell,BellFitz}, or other related work as reported~\cite{Nelson15} suggests that he viewed a small acceleration to reach beyond Einstein\rq s SR.

Such thinking  is not accurate as understanding of forces and acceleration is an essential component in the formulation of foundational theories. We recall that GR was born out of the effort to understand the force we call gravity. However, acceleration due to gravity alone does not exist by virtue of the equivalence of gravitational and inertial mass. In GR, a sub-domain of the theory of relativity, gravity is accommodated in terms of space-time modification by material bodies. Point particles move on generalized straight lines, geodetics. A similar interpretation was attempted by Kaluza~\cite{MBlago2002} for the electromagnetic (EM) force, introducing a 5th dimension.  To this day a fully consistent understanding of EM force at a level that rivals GR has not been found. Therefore when we introduce forces and acceleration, created in an EM context, this is done in an intuitive-empirical \lq Lorentz\rq\ approach.

To set a body in motion, a force -- and thus acceleration -- is applied. However, as just noted we do not know  acceleration exists. We must therefore ask, how does the concerned body know it is accelerated? Or maybe a body cannot feel an acceleration~\cite{Moskal}? I believe in general acceleration must be felt by a body in order to justify creation of radiation; that is, body detached propagating fields. From an observer\rq s viewpoint, without acceleration there would be no way to explain how it is possible to move dynamically a body from one inertial frame of reference to another. This dynamical change process will allow us to perform a laboratory experiment to measure the Lorentz-Fitz\allowbreak Gerald body contraction, which as we will explain has so far not been experimentally observed.
 
{\bf Body contraction.} --- Let us look at the Lorentz-Fitz\allowbreak Gerald body contraction using the Lorentz-Bell approach. We consider an object at rest with respect to us (laboratory) and measure its body length with an instrument in this laboratory frame. The measurement of the entire object is carried out at equal laboratory time. This is an important condition since simultaneity of any two events is present in one reference frame only. It is natural to measure at equal time in laboratory, and this assures the unique definition of the measurement of body size, in particular relevant if and when the body is not at rest in the laboratory frame. 

We set the object of interest slowly in motion. We continue to size-up the body at equal time in the laboratory frame. The object we have set in motion is observed to be contracted in its body length parallel to body motion direction by the Lorentz-factor $\gamma=1/\sqrt{1-\beta^2},\ \beta=v/c $. Using the Einstein pedagogy this is seen evaluating with the help of Lorentz transformation the distance between the two ends of the moving body at equal time in the laboratory frame -- a measurement at equal time in the body  co-moving reference frame would of course produce no body contraction, see Einstein 1911 \loccit. Comparing to the measurement of momentum, or the kinetic energy of a moving body, we find the situation as noted by Rindler 1960 \loccit.

We now consider the \lq\lq two rocket\rq\rq\ example championed by Bell~\cite{JSBell}. The two rockets are independent in their individual motions. However, what they do is synchronized by the laboratory observer. They are very slowly accelerated in an identical way with respect to the laboratory. The material object of interest to us is now a very thin rod placed between these two idealized rockets. 

There are two different \lq lengths\rq\ of interest: the spatial distance between the rockets, and the material size of the moving rod. Integration of time dependent motion of the two rockets shows that this spatial distance always remains constant, no matter how fast these rockets ultimately move, a result seen in Ref.\,\cite{JSBell} -- space and time are unaffected by material bodies in absence of gravity. On the other hand, the material rod placed between the two rockets is Lorentz-Fitz\allowbreak Gerald body contracted. We see the gap open since the rod is contracted, but the spatial distance between the rockets is not.

The physical effect, \ie\ the rod does not anymore connect the two rockets also arises in Einstein\rq s pedagogy. To see this we Lorentz-transform the laboratory observer on to one now inertially moving rocket to which the rod was attached. For this rod co-moving observer, the rod is the same length as before. However, this observer sees the other rocket further away than the length of the rod, since the time simultaneity established in the original laboratory frame does not apply in the rod reference frame. 

Here is the Einstein pedagogical mind-trap: nothing happened to space-time in the process of Lorentz transforming the lab observer to sit on one of the rockets, and yet the rod is not connecting the rockets. To some this means that space is somehow affected by body motion. A century after Einstein this misrepresentation of SR is found in many introductory books. This creates need for physical reality approach prompting J\' anossy\rq s~\cite{Janossy71} book; Bell\rq s paper~\cite{JSBell}; and induced in my mentor Walter Greiner interest in helping to create our classic text (in German)~\cite{Drelativ}. A followup in English is today available~\cite{Rafelski:2017hyt}.

{\bf Time dilation, body contraction and Michelson-Morley experiment.} --- To recapitulate: a) the body contraction measurement is carried out at equal time in the observer\rq s frame of reference; and (as all are familiar with) b) the time dilation is recorded by a proper body clock co-moving with respect to the body. Once the above procedure of measurement is defined, the outcome of measurement of both body size and proper time is unique. The fact that both are governed by the Lorentz-factor $\gamma$ is not evidence that these effects are the same, or that one somehow explains the other  as one often sees incorrectly stated in SR introductions. 

Hereto, there is a reference position inversion: body contraction relates to equal in time measurement by a typically laboratory resting observer. However, time dilation relates to the proper body clock, at rest in the moving body, thus referring to equal in space measurement in the (moving) body reference frame. Both situations introduce unique and vastly different measurement procedures. In principle the two measurements are entirely unrelated; we cannot use one to argue for the other.
 
These two unrelated procedures can complement each other, for example to build a light path clock that scores proper time independent of the clock orientation. In such a clock, a light pulse is bouncing between two mirrors attached firmly to a solid material base. A time tick on this light clock is the return time of the signal. \\
\indent {\bf a)} Considering the transverse to direction of motion light path orientation, we find that the optical path chasing the moving mirror turns ever longer as the speed of the clock increases. The lengthening of the optical path in the vacuum is required so that there is time dilation; the clock with the longer optical path scores fewer \lq\lq ticks\rq\rq\ compared to a clock at rest in laboratory . \\
\indent {\bf b)} Consider the optical path of the light clock with the mirror axis rotated in any other direction: we find the same answer as in the case of \lq\lq transverse light clock\rq\rq\ only if the material body providing the base for the mirrors is subject to the Lorentz-Fitz\allowbreak Gerald body contraction in the direction of motion.

The orientation independence of the mirrored optical path with mirrors attached to a moving body is what assures that the Michelson-Morley (MM) experiment has a null outcome. The Lorentz-Fitz\allowbreak Gerald body contraction is \lq correcting\rq\ the optical path effect of moving mirrors. Some argue that this constitutes measurement of body contraction in a co-moving reference frame. Einstein when he assures that the Lorentz-Fitz\allowbreak Gerald body contraction is real~\cite{AEEhrenfest} argues that a co-moving observer cannot detect a Lorentz-Fitz\allowbreak Gerald body contraction. 

I interpret this that in Einstein\rq s eyes the unobservability of absolute motion by the MM experiment (and thus the proper functioning of the light clock) is due to Galileo's relativity principle forbidding the measurement of absolute motion, and this principle cannot be used in circular fashion to claim body contraction in the direction of motion to explain the null result of the MM experiment. Adopting this point of view, we are losing the one and only way that so far  could be used to claim experimental verification of the Lorentz-Fitz\allowbreak Gerald body contraction.
 
{\bf Accelerated motion} --- An experiment directly measuring the Lorentz-Fitz\allowbreak Gerald  body contraction has never been carried out. The experiment requires within a laboratory sized setup, taking an extended body from rest to ultra-relativistic speed and the measurement of the contracted body size in the direction of motion. To achieve these goals we will use the largest force available to impart a relativistic speed  on a small material body, and we will probe the body with a laser beam.

However, how do we know a body and not the observer is accelerated? We recall Mach\rq s idea of introducing the fixed star reference frame. The existence of this in-principle way to distinguish between inertial and accelerated motion solves only part of the problem: we need to know who is accelerated instantaneously; fixed stars are too far away to be useful beacons. It must be that the empty space-time itself provides us with the information allowing the recognition of acceleration. 
 
Einstein reintroduced in 1919/20 non-material \ae ther filling all space-time. Einstein argues (original German in reference)~\cite{AEaether} \lq\lq To summarize we can say that according to the general theory of relativity, space is endowed with physical qualities; in this sense the \ae ther exists. According to the general theory of relativity, space without \ae ther is unthinkable: without \ae ther light could not only not propagate, but also there could be no measuring rods and clocks, resulting in nonexistence of space-time distance as a physical concept. On the other hand, this \ae ther cannot be thought to possess properties characteristic of ponderable matter, such as having parts trackable in time. Motion cannot be inherent to the \ae ther.\rq\rq\ 

In the context of modern quantum field theory we instead introduce the structured quantum vacuum, which fulfills the role that Einstein assigned to the non-material \ae ther. This is not the place to dwell in depth on how the Higgs field provides mass to particles, or how the non-perturbative chromodynamic quark-gluon vacuum fluctuations generate the dominant fraction of the mass of matter. For the purpose we pursue here, it suffices to accept that the structured quantum vacuum provides matter with inertial mass. In this sense the quantum vacuum is Einstein\rq s 1919/20 non-material \ae ther. 

While we do not have an explicit and fundamental formulation of how material bodies know they are accelerated, the way the modern quantum field theory provides inertial mass that generates resistance to force, material bodies evidently know if their state of motion is inertial or not. Any particle, including the electron, clearly knows about being accelerated relative to the quantum vacuum state that defines the class of inertial observers.   

From this consideration it follows that the principle of relativity does not apply in the presence of acceleration; there is no equivalence between the case of an accelerated body observed by inertial observer with the case of an inertial body looked at by an accelerated observer irrespective of the magnitude of acceleration. Our argument does not apply to gravity, which would not create an acceleration but motion in curved space-time. 

{\bf Body contraction experiment.} --- To accomplish our goal to build a laboratory-sized experiment we consider an ultra-intense ultra-short laser pulse shot at a thin (micron) foil. Such a pulse in its focal point can act  as a micron-sized hammer pushing out of the foil an electron cloud accelerated to ultrarelativistic motion with a high value of Lorentz-factor $\gamma_e$. The emerging electron cloud compared to the original foil thickness  will be Lorentz-Fitz\allowbreak Gerald compressed by  $\gamma_e$.

This situation is reminiscent of the relativistic atomic nuclei being accelerated at the giant collider LHC to $\gamma_\mathrm{HI}\simeq 5000$, but in the present case we use a giant laser to produce a pulse serving to accelerate over a micron scale distance an electron cloud to ultrarelativistic speed. Values of $\gamma_e\sim \gamma_\mathrm{HI}$ could become available in the near future at Extreme Light Infrastructure (ELI) facilities in Europe. 

A moving electron cloud acts as a relativistic mirror for a low intensity laser light bounce. The capability of the ultrarelativistic mirror to function depends on the electron cloud density; laser light can scatter coherently from a sufficiently high density cloud -- what is low and high density is determined by comparing mean electron separation to the light wavelength. 

Many believe that such an electron cloud based relativistic mirror holds great promise to generate cheaply a relatively compact coherent X-ray light source~\cite{Boulanov,Kiefer}: the visible light in coherent scattering should be  stepped up in energy by a factor $\propto\gamma^2$ as was already pointed out by Einstein in the 1905 SR paper~\cite{EinsteinSR}, see also Section 19.3 of Ref.~\cite{Rafelski:2017hyt}. The quadratic Lorentz-factor arises from two Lorentz transforms, first into the rest-frame of the mirror, and upon reversal of the propagation direction of the light motion, transform back to the laboratory frame.

The condition for coherent upscale of the reflected light beam into the X-ray regime is satisfied for 
$$
\gamma_e^\mathrm{max}\lesssim (\lambda/d_e)^a\,,\quad a\simeq 1/3\,,
$$
where $d_e$ is electron mean separation as measured within the moving mirror cloud. The value of the power index $a$ relates to the fact that only one of three body sizes of the electron cloud is contracted. Both the probing laser wavelength $\lambda$ of the back scattered light and $d_e$ can be varied in an experiment. Variation of $d_e$ arises from temporal dilution due to electron Coulomb repulsion-explosion. We note that establishing $\gamma_e^\mathrm{max} \propto \lambda^{1/3}$ will suffice to demonstrate that the electron density in the moving cloud was created with a Lorentz-Fitz\allowbreak Gerald body contraction.

It is important to realize that the reinterpretation of the electron cloud body contraction as an effect of the probing laser beam Doppler wavelength reduction is incorrect. As was recalled recently~\cite{Margar2017}, the Doppler effect is created in the process of \lq observation\rq\ of the Lorentz invariant light wave phase, thus in the present case in the process of collision of the Lorentz-Fitz\allowbreak Gerald body contracted electron cloud with the incoming coherent and invariant light phase.

{\bf Conclusions.} --- Einstein in 1905 sidestepped acceleration because it required much further study and development. The creation of GR shows by example that it was possible to resolve and understand a force, that of gravity, in a fundamental way. However, other fundamental forces were not accommodated, what we call SR is an incomplete theoretical framework. The only reason that we rarely notice this is that the current physical domain encompasses in most extreme but conventional laboratory conditions nano-strength acceleration~\cite{Rafelski:2017hyt}.

Although the introduction of the Lorentz-Fitz\allowbreak Gerald body contraction was prompted by the Michelson-Morley null result, seen through Einstein\rq s eyes this null result established the principle of relativity that forbids measurement of absolute motion. Adopting this principle Einstein obtained the key results of SR including Lorentz-Fitz\allowbreak Gerald body contraction of a moving body. But he did not see in circular fashion the MM experiment as providing evidence for  body contraction.

In the coming decade(s) the forthcoming accessibility of high acceleration experiments should pave the way to accommodating acceleration at a fundamental level in SR. This should occur in parallel to a better understanding of Einstein\rq s \ae ther, \aka\ the structured quantum vacuum, which allows us to distinguish inertial and non-inertial bodies and observers. 

Among the forthcoming high intensity light pulse acceleration experiments the interaction of a laser beam of modest intensity with a flying relativistic electron mirror can serve to generate a reflected coherent beam in the X-ray energy range. We have pointed out that in such experiments the Lorentz-Fitz\allowbreak Gerald body contraction that the electron cloud can experience at the time of formation enhances the range of coherent back-scattering and at the same time offers the opportunity to experimentally explore the Lorentz-Fitz\allowbreak Gerald body contraction.


\vfill\eject
\end{document}